\documentclass{article}

\usepackage[final]{neurips_2026}

\usepackage[utf8]{inputenc}
\usepackage[T1]{fontenc}
\usepackage{hyperref}
\hypersetup{hidelinks}  
\usepackage{url}
\usepackage{booktabs}
\usepackage{amsfonts}
\usepackage{amsmath}
\usepackage{nicefrac}
\usepackage{microtype}
\usepackage{graphicx}
\usepackage{xcolor}
\usepackage{siunitx}
\usepackage{placeins}   
\usepackage{flafter}    


\makeatletter
\AtBeginDocument{\def\@notice{}}
\makeatother

\title{Amortized Opacity Marginalization Improves C/O\\
Interval Calibration for Brown-Dwarf Retrievals}

\author{%
  Kellen Heraty \\
  sculptorai.org \\
}

\begin{document}
\maketitle

\begin{abstract}
Molecular line lists disagree at a level that measurably shifts retrieved C/O,
so retrievals conditioned on a single list omit a dominant systematic from their
error bars. Classical marginalization re-runs a full retrieval under $K$ opacity
realizations per object. We move that cost into simulation: while generating the
training set we randomize opacities (smooth multiplicative Fourier fields in
$\log\nu$ plus a discrete line-list mixture, under a deliberately broad prior
informed by measured inter-list differences), so a neural posterior estimator
learns an already-marginalized posterior. Conceptually, this is the amortized
analogue of repeating a conventional retrieval under many plausible opacity
realizations and averaging the resulting posteriors. On perturbed held-out spectra
($n=2048$) a standard NPE covers nominal 90\% C/O intervals 43.3\% of the time
against 82.4\% for a compute-matched marginalized network; three-seed control
ensembles reach 93.2\%
on clean data but 61.4\% under perturbation, against 89.4\% marginalized,
isolating the deficit as opacity-induced (TARP joint expected-coverage deviation
0.005 against 0.035 for a single control network). An axis ablation traces the gain to the continuous field
(45.4\% for the discrete swaps alone, 79.3\% for the field alone). In simulation, real
JWST bad-pixel masks with median filling cut the marginalized ensemble's C/O
coverage from 0.894 to 0.683; an ensemble retrained with mask augmentation holds
0.918 under the full deployment preprocessing ($n=1024$). We apply
the networks to one pipeline-combined JWST/NIRSpec G395H spectrum each of 14
late-T and Y dwarfs. The recommended mask-augmented ensemble returns C/O medians between
0.24 and 0.62 and contains the Hood et al.\ (2024) value of 0.53 for the T8
benchmark 2MASS J0415$-$0935, which every fixed-opacity control network excludes.
With 90\% intervals spanning about 0.43 of the prior width, this is a
consistency check, not a validation.
\end{abstract}

\section{Introduction}
\label{sec:intro}

Atmospheric retrieval infers composition (for instance the carbon-to-oxygen
ratio C/O) and thermal structure from a spectrum by comparing it to a radiative
forward model whose molecular opacities come from line lists
(HITEMP, ExoMol, and others) that disagree: in the ExoJAX retrievals of
VLT/CRIRES spectra of Luhman\,16AB by \citet{yama2026}, line-list choice is the
dominant systematic on C/O at the $\sim$7\% level. It joins a long line of retrieval systematics: broadening choices
\citep{gharibnezhad2019}, temperature-profile parameterization
\citep{rocchetto2016}, inter-code differences \citep{barstow2020}.
Error-inflation devices, a fitted ``$10^{b}$'' flux
jitter \citep{line2015} or a Gaussian-process noise term, absorb some unmodeled
variance, but they inflate a smooth, parameter-independent noise budget and
cannot restore per-parameter calibration against a structured, parameter-coupled
opacity error. The principled remedy, marginalizing over
line-list choice, classically propagates opacity uncertainty through $K$
complete retrievals per object, as \citet{niraula2022,niraula2023} did for
transmission spectra: correct, but expensive and repeated per target.

\paragraph{Contribution.}
We amortize opacity marginalization. Instead of $K$ retrievals per object we pay
once, in simulation, by randomizing opacities while generating the training set
for a neural posterior estimator (NPE)
\citep{papamakarios2016,greenberg2019,cranmer2020}, so the resulting posterior is
marginalized over the specified line-list-uncertainty prior by construction and
evaluates in a single amortized forward pass. This is, to our knowledge, the
first controlled calibration-\emph{width} study for brown-dwarf emission
retrieval under line-list uncertainty, against un-marginalized and
recalibrated-control baselines. Classical marginalization is
tractable; the contribution is its cost structure and calibration. The aim is not
tighter constraints. It is a quoted uncertainty that remains meaningful when the
opacity model is uncertain. All claims
are conditional on the prior of \S\ref{sec:method} and scoped to central-interval
C/O coverage (\S\ref{sec:limits}).

\paragraph{Related work.}
Simulation-based inference with normalizing flows
\citep{papamakarios2021,cranmer2020} underlies neural retrieval, and NPE has
been applied to exoplanet spectra with opacities held fixed
\citep{marquez2018,vasist2023}. A parallel literature shows amortized posteriors are
frequently \emph{overconfident} unless controlled \citep{hermans2022,%
delaunoy2022}, and develops misspecification-aware NPE
\citep{ward2022,cannon2022}, deep ensembling \citep{lakshminarayanan2017}, and
coverage diagnostics beyond marginal simulation-based calibration (SBC), notably
TARP \citep{lemos2023,talts2018}. We combine these threads: opacity randomized inside the simulator after
\citet{tobin2017}, and ensembling against seed-level under-dispersion.

\section{Method}
\label{sec:method}

\paragraph{Forward model.}
We use the ExoJAX-based \citep{kawahara2022,kawahara2025} SCULPTOR forward model for
brown-dwarf emission over JWST/NIRSpec G395H ($1850$--$3550$\,\si{\per\cm},
$\approx$2.8--5.4\,\si{\micro\meter}), with 15 retrieved parameters: C/O,
metallicity [M/H], a five-knot temperature profile ($T_1$--$T_5$), radius, gravity
$\log_{10}(g/\mathrm{cm\,s^{-2}})$, vertical mixing
$\log_{10}(k_{zz}/\mathrm{cm^{2}\,s^{-1}})$, a two-parameter
($f_{\mathrm{sed}},\sigma_c$) sedimentation cloud after \citet{ackerman2001} in
its grey, geometric-optics limit, and radial/rotational/calibration nuisances
(ranges and abundance recipe in Appendix~\ref{app:repro}). \emph{C/O is the
elemental carbon-to-oxygen number ratio input to equilibrium chemistry}
(pyFastChem, \citealt{stock2018,stock2022}; \citealt{asplund2009} solar scale,
``scale-C''), post-processed with CO/NH$_3$ quenching set by $k_{zz}$. With no
explicit oxygen-condensation correction it is a bulk elemental ratio, not a
gas-phase-after-rain-out value.

\paragraph{Perturbation model.}
For each simulated spectrum we perturb the opacity of every major species by a
smooth multiplicative field,
$\kappa(\nu)\!\to\!\kappa(\nu)\exp\!\big(c_0+\sum_{m=1}^{4}
[a_m\cos(2\pi m u)+b_m\sin(2\pi m u)]\big)$, a 4-mode Fourier expansion in
normalized $\log$-wavenumber $u$, \emph{and} we draw the discrete line list from
a mixture. The nine coefficients per species are drawn from zero-mean Gaussians
with per-species, per-mode standard deviations (Table~\ref{tab:opamp}),
independently across species and draws. All six G395H species carry the continuous field, so all four C- and O-bearing
carriers of C/O are perturbed; the discrete swap is available only for H$_2$O (POKAZATEL,
\citealt{polyansky2018}, against HITEMP) and CH$_4$ (HITEMP,
\citealt{hargreaves2020}, against ExoMol-MM, \citealt{yurchenko2024}), whose four
combinations cycle uniformly over shards. The Fourier field is not intended as a
literal physical model of line-list errors. It provides a continuous family of
plausible opacity distortions rather than only a few fixed opacity tables that the
network may memorize, the distinction the one-axis ablation of
\S\ref{sec:results}(h) isolates. Amplitudes come from
HITEMP-vs-ExoMol cross-section ratio spectra measured over a photospheric
$(T,P)$ grid and sit above those discrepancies (Appendix~\ref{app:repro}):
over-dispersion costs width, under-dispersion costs coverage, and we choose
width. Two caveats. The prior is a \emph{modeling
choice informed by}
measurements, since \citet{yama2026} measure a C/O \emph{shift}, not a
fractional opacity error. Swap and field also derive from the same inter-list
differences, so they partly double-count; we read the field as within-list
structural error, the band-strength and line-shape residuals four fixed swap
realizations cannot express. The amplitude sweep scales only the field.

\paragraph{Datasets and NPE.}
On a TPU v4-64 we generated two matched 1M-spectrum sets (7{,}813 shards each),
perturbed and clean, identical otherwise. The NPE is a bounded neural spline flow
with a 1D-CNN embedding, trained with best-validation checkpointing
(Appendix~\ref{app:repro}); the marginalized model is trained on perturbed data
(three seeds), the control on clean data, with identical architecture and flags.
Opacity coefficients are recorded but dropped from the label set
\citep{alsing2019}.

\section{Results}
\label{sec:results}

All coverages are marginal central-interval coverages, the fraction of held-out
truths inside the posterior central interval at the stated level; ``overall'' is
the unweighted mean over the 15 parameters, \emph{not} simultaneous coverage (protocols in Appendix~\ref{app:repro}).

\paragraph{(a) Coverage under perturbation, and scaling.}
The key controlled comparison uses three-seed ensembles with identical
architecture, training budget and pooling: the control ensemble covers C/O at
$0.932$ on a clean held-out set ($n=512$) but $0.614$ on the perturbed
factorial split ($n=2048$), where the marginalized ensemble reaches $0.894$.
The deficit is opacity-induced, not an ensembling effect, and marginalization
removes most of it.
Table~\ref{tab:coverage} repeats it at single-network scale ($0.433$ against
$0.824$), all gaps tens of binomial standard errors. In other words, when the
fixed-opacity network is applied to spectra affected by opacity uncertainty, an
interval labeled as 90\% contains the true C/O less than half the time. A single
control network reaches only $0.62$ even on clean data, so part of that advantage
is generic augmentation. Nor is it averaging over broken regimes: raw C/O coverage holds in every $T_4$
and SNR quintile where the control's collapses (Appendix~\ref{app:repro}). From 800 to 3{,}000 training shards the
control's C/O coverage \emph{drops} $0.57\to0.47$ ($n=512$ perturbed
protocol): more data sharpens the posterior
around a misspecified forward model instead of correcting it.

\begin{figure}[h]
\centering
\includegraphics[width=\linewidth]{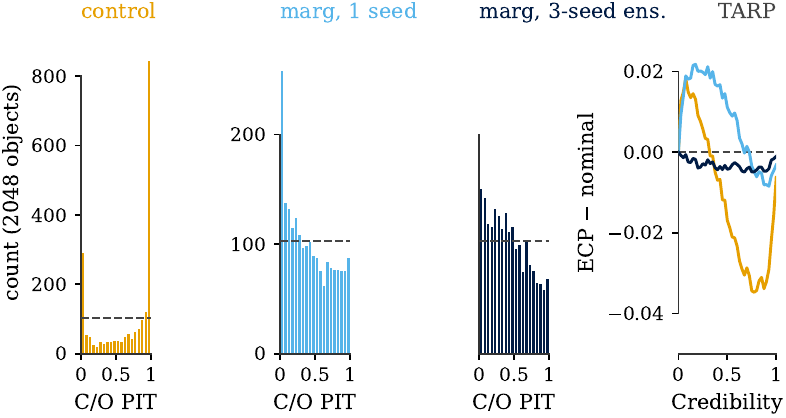}
\caption{Simulation-based calibration on the $n=2048$ split, 400 posterior
samples per seed. Left three panels: C/O PIT histograms, where a calibrated
posterior is uniform (dashed); the two marginalized panels share a second,
expanded vertical scale. All $\chi^2$ quoted here are for the C/O PIT histogram
alone (20 bins, 19 dof), not the 15-parameter pooled statistic of
\S\ref{sec:results}(b). The control is strongly U-shaped (overconfident, rank
mean 0.673, C/O $\chi^2=6315$); a single marginalized seed is closer to uniform
(rank mean 0.416, C/O $\chi^2=328$); the three-seed ensemble is closer still but
retains a residual asymmetric bias (rank mean 0.427, C/O $\chi^2=146$). Right panel: TARP
expected-coverage probability minus nominal (50 reference points), where zero is
exact joint calibration; the three curves take the colours of the PIT panel
titles. The ensemble's maximum deviation is 0.0049, against 0.0218 for a single
seed and 0.0348 for the control.}
\label{fig:sbc}
\end{figure}

\paragraph{(b) The three-seed ensemble reaches near-nominal coverage.}
A single marginalized network plateaus in overall coverage near 0.83, matching
the flow under-dispersion ensembling corrects \citep[cf.][]{hermans2022};
pooling one, two, three seeds gives C/O $0.844$, $0.842$, $0.893$, so seed-
count saturation is untested. Joint
calibration, measured by TARP \citep{lemos2023}, is about sevenfold tighter
for the marginalized ensemble than for a single control network: maximum
expected-coverage deviation $0.0049$ (bootstrap 95\% $[0.0025,0.0144]$) against
$0.0348$ (Fig.~\ref{fig:sbc}). A single marginalized network reaches $0.0218$
(bootstrap 95\% $[0.0167,0.0280]$), so ensembling accounts for most of that gap,
and the comparison is not between equal numbers of networks. SBC still finds residual
miscalibration: pooling the probability-integral-transform (PIT) values of all 15 parameters
($30{,}720$) the ensemble rejects uniformity
($\chi^2_{\mathrm{pooled}}=116.2$, 19 dof, $p<10^{-14}$), and the C/O histogram is
worse than the pool ($\chi^2_{\mathrm{C/O}}=146$, rank mean $0.427$;
Fig.~\ref{fig:sbc}), a location or shape bias that central coverage does not
reveal, so we do not claim full joint calibration.

\paragraph{(c) Recalibration is not a substitute.}
A cheaper baseline leaves opacities fixed and recalibrates the control post hoc
with the per-parameter monotone PIT-CDF map of \citet{kuleshov2018}, fit on a disjoint calibration half; being
asymmetric it
corrects dispersion \emph{and} location, a stronger baseline than width
inflation. On the leakage-free split (2{,}048 calibration and 2{,}048 test
objects, frozen maps), C/O coverage at nominal 0.90 goes
$0.433/0.705/0.824/0.881$ for raw control, recalibrated control, raw
marginalized and recalibrated marginalized, aggregate coverage $0.747/0.869/0.831/0.894$:
recalibration recovers \emph{aggregate} coverage but not C/O, since a frozen
marginal map cannot fix a per-object conditional bias (quintiles in
Appendix~\ref{app:repro}). In the $2\times2$ ensemble
factorial the strongest control stack
(recalibrated control ensemble, 0.830 at width/prior 0.20) is beaten on both
counts by the \emph{raw} marginalized ensemble (0.894 at 0.17), and
recalibrating both sides keeps the advantage (paired bootstrap
$\Delta$coverage $+0.058$ at $\Delta$width $-0.048$;
Appendix~\ref{app:repro}), though
it slightly lowers the near-calibrated ensemble
($0.894\to0.879$). This rules out this recalibrator, not every post-hoc scheme.

\paragraph{(d,e) Prior amplitude and held-out discrete cache.}
Coverage falls monotonically as the prior is misspecified, from 0.94 at
$0.43\times$ the trained amplitude to 0.75 at $2.14\times$, so the claim does
not extend to $2\times$. A model retrained with the CH$_4$-ExoMol cache withheld
covers never-seen shards from it as well as matched in-cache data (0.821 against 0.790), a stability result on a test that is not family-blind,
only the discrete cache being withheld (Appendix~\ref{app:repro}).

\begin{figure}[h]
\centering
\includegraphics[width=3.05in]{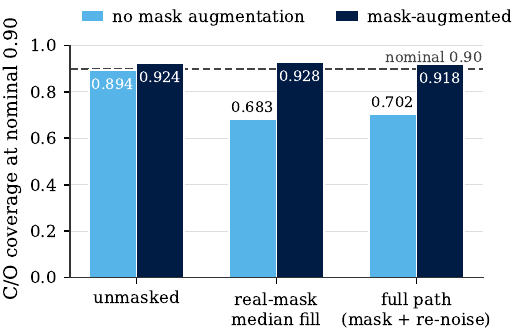}
\caption{C/O coverage at nominal 0.90 on $1{,}024$ perturbed held-out objects for
three-seed marginalized ensembles trained without (light) and with (dark) mask
augmentation, scored on the same spectra. The masks come from the 14 real
spectra of \S\ref{sec:results}(g) (median good-pixel fraction 0.89). Masking
costs the un-augmented ensemble about a quarter of its coverage; the
mask-augmented ensemble stays above nominal.}
\label{fig:deploy}
\end{figure}

\paragraph{(f) Deployment preprocessing costs calibration; mask-augmented
ensembling repairs it.}
On real spectra the deployment path masks bad pixels and the NRS1/NRS2 detector
gap, median-fills them and conditions on the median SNR, none of which the
training set saw. To measure the cost, we apply the masks of the 14 real spectra
of (g) (median good-pixel fraction 0.89) to $1{,}024$ perturbed held-out
simulations, whose truths are known, and score every network on the same masked
spectra (Table~\ref{tab:stress}, Fig.~\ref{fig:deploy}). Masking and median
filling cut the marginalized ensemble's C/O coverage from 0.894 to 0.683, and
the full path, which adds flat re-noising at the median SNR, gives 0.702, so the
masking does the damage. Three networks retrained with mask augmentation and
pooled hold 0.924 unmasked, 0.928 masked and 0.918 on the full path, where
their overall coverage is 0.894. The fix costs width: the mask-augmented
ensemble's median 90\% C/O interval is 0.209 of the prior width unmasked, against
0.164 for the un-augmented ensemble. Two limits apply. The stress set is
simulated, and the mask-augmented networks were trained on masks drawn from the
105 products of versions 1 and 2, which include these 14, so this is not a test
on unseen mask geometries. Versions 1 and 2
reported a far larger collapse, from 0.901 to 0.479, because most of their 105
masks came from single-detector files that cover about half the band.

\begin{figure}[h]
\centering
\includegraphics[width=\linewidth]{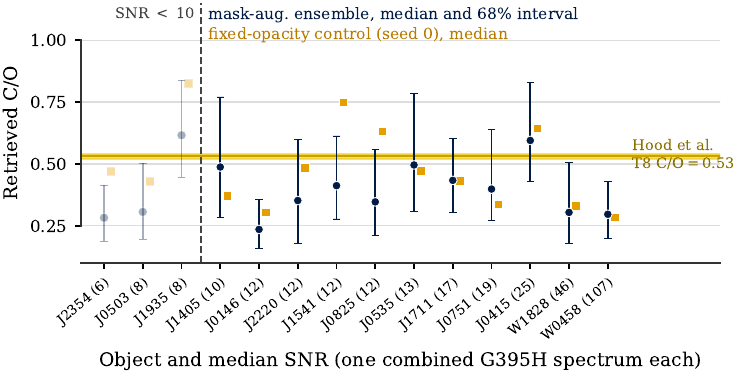}
\caption{Retrieved C/O for the 14 objects of Table~\ref{tab:sample}, one
combined G395H spectrum each, ordered by median SNR (in parentheses). Navy: the
mask-augmented three-seed ensemble, median and 68\% interval. Orange: median of
the fixed-opacity control, seed 0. Gold: the \citet{hood2024} value for the T8
benchmark J0415. Objects left of the dashed rule fall below the training SNR
range and are drawn faded.}
\label{fig:real}
\end{figure}

\paragraph{(g) Application to real JWST spectra.}
The sample is one pipeline-combined (level-3) NIRSpec G395H/F290LP spectrum per
object for 14 late-T and Y dwarfs from programs 1189 and 2124
(Table~\ref{tab:sample}, Fig.~\ref{fig:real}). Versions 1 and 2 of this paper
used 105 products of these programs, which mixed the combined spectra with
single-exposure, single-detector files and so counted most objects many times,
often as half-band spectra. Seven of those products held no source: three
fixed slits of the 1189 observation that did not contain WISE~1828$+$2650, and a
failed 2124 visit to J1935$-$1546 (its combined spectrum and three exposures),
which the program repeated. This version drops all of them. Each spectrum is
resampled onto the model grid, its bad pixels and the NRS1/NRS2 detector gap are
masked (good-pixel fraction 0.89 for the slit spectra, 0.93 for the IFU
spectrum) and median-filled, and the network is conditioned on its median SNR.
Eleven objects fall inside the training SNR range of $[10,200]$; the other three
(SNR 6.3 to 8.1) are shown in Fig.~\ref{fig:real} but left out of the summaries
below.

On the 11 in-support objects the recommended deployment, the three
mask-augmented seeds of (f) pooled, returns C/O medians from 0.24 to 0.59 and a
median 90\% interval of 0.43 of the prior width. The mask-augmented seeds also
agree more closely with each other: the pairwise median $|\Delta\mathrm{C/O}|$ between seeds is
0.09 for the mask-augmented networks, against 0.25 for the un-augmented
marginalized networks and 0.23 for the controls. For the T8 benchmark
2MASS J04151954$-$0935066 (SNR 24.5), \citet{hood2024} report C/O
$=0.53\pm0.01$, a statistical bar only. The mask-augmented ensemble gives
$0.595$, 68\% interval $[0.430,0.829]$, and each of its three seeds contains
$0.53$ at 68\% and 90\% (Table~\ref{tab:real}). Every control network misses it
at both levels, with medians of 0.64 to 0.70 and 90\% intervals that stop at
0.54 or above. The un-augmented marginalized seeds split: two contain $0.53$ at
both levels and the third, with median 0.92, at neither. One object with an
interval this wide is weak evidence. Hood's oxygen convention matches ours, but
their chemistry, wavelength coverage and forward model differ, so this is a
consistency check (Appendix~\ref{app:repro}).

\paragraph{(h) The continuous field carries most of the calibration benefit.}
Two networks matched to marginalized seed 0 in seed, schedule and shard count
vary one perturbation axis (Table~\ref{tab:ablation}). Trained on the
discrete swaps alone, a network is no better calibrated than the fixed-opacity
control ($0.454$ raw against $0.433$).
Trained on the field alone it reaches $0.793$, close to the
$0.825$ of the full mixture. Both axes still bite: on spectra perturbed one axis
at a time the control covers C/O at $0.614$ and $0.472$ against $0.433$ when
both act, while the marginalized ensemble, not retrained, holds $0.963$ and
$0.910$.

\FloatBarrier
\section{Limitations}
\label{sec:limits}

All results are conditional on the prior of \S\ref{sec:method}, whose two
caveats carry over, and the fields are $T/P$-independent whereas real line-list
error is not. Coverage is marginal and PIT uniformity is still rejected (b), so
we scope the claim to central C/O intervals. Simulator failures induce a parameter-dependent acceptance function,
especially in C/O, so calibration is conditional on successful generation, not
on the nominal prior.
The mask-augmented networks were trained with masks drawn from the 105-product
set of versions 1 and 2 (median good-pixel fraction 0.47, most of them
single-detector half-band masks) and are tested here with the 14 clean masks, so
masks from other programs and reductions remain untested. Real-data
preprocessing stays coarse, and no posterior-predictive check is reported for
the clean sample. Marginalization widens intervals: on the simulated split the
marginalized network's median 90\% C/O interval is $2.6\times$ the control's
for single networks and $2.1\times$ for the ensembles
(Table~\ref{tab:coverage}). Posterior fidelity against a classical NUTS
retrieval is not tested. No line-shape lever we tried (LSF nulls, broadening rescalings, $\chi$-factor
CO$_2$ line mixing, \citealt{perrin1989}) recovered the $\sim$7\% floor, which
is why we marginalize over it.
The Fourier field also stands in for an error with a measurable origin: the
three CO configurations of \citet{yama2026} differ only in pressure broadening,
and their $\sim$7\% C/O spread is almost entirely air- against H$_2$-broadened
HITEMP. Perturbing the saved opacity
grids at that scale (half-widths $\times1.30$, temperature exponents $-0.15$,
all six species) leaves the marginalized
ensemble's C/O coverage at $0.901$ against $0.910$ on the matched reference set,
inside the $0.014$ binomial standard error. Overall coverage falls
$0.906\to0.858$, driven by the temperature knots, worst at $T_3$ and next at
$T_2$, with $\log g$ behind them; widths move by at most 7\%, so the mismatch
displaces the posterior rather than shrinking it (Table~\ref{tab:stress}). At this amplitude and sign, composition survives a misspecification the prior
never contained and thermal structure does not, though the C/O interval spans
$0.18$ of the prior width, so some of that survival is margin rather than generalization. We read the
damage as landing on the parameters setting photospheric pressure and hence
line width, a mechanism this test does not isolate and a broadening prior anchored in the
laboratory bounds of \citet{hosokawa2025} would target
(Appendix~\ref{app:repro}).

\FloatBarrier

\bibliographystyle{plainnat}
\bibliography{references}

\section*{Data, code and acknowledgments}
Code, result JSONs and the paper source are at
\url{https://github.com/kaileh57/opacity-marginalization}; datasets, trained
weights, priors and held-out evaluation sets are at
\url{https://huggingface.co/datasets/Kaileh57/opacity-marginalization}.
We thank Hajime Kawahara for ExoJAX and for his guidance throughout this
project, from its earliest framing to the experiments reported here.
Supported with Cloud TPUs from Google's TPU Research Cloud.

\textbf{Use of generative AI.} The author used Claude (Anthropic) to help
with LaTeX formatting and to edit the manuscript text for clarity. Claude was
not used to generate data, run calculations, produce analysis or select
references. The author reviewed every edit and takes full responsibility for
the content. To validate the final text, every number reported in the paper
was checked against the result files released with the code, and every
reference was checked against its arXiv or journal record (this check
corrected one page range).

\appendix
\section{Reproducibility}
\label{app:repro}

\textbf{Parameters and priors (uniform).} $T_1\!\in\![150,1500]$,
$T_2\!\in\![200,2000]$, $T_3\!\in\![300,2500]$, $T_4\!\in\![500,3000]$,
$T_5\!\in\![800,3500]$\,K (temperature knots at
$\log_{10}(P/\mathrm{bar})=-5,-3,-1,1,3$); [M/H]\,$\in[-1,1.5]$\,dex;
C/O\,$\in[0.1,1.5]$; $\log_{10}(k_{zz}/\mathrm{cm^{2}\,s^{-1}})\!\in\![6,10]$;
$\log_{10}(g/\mathrm{cm\,s^{-2}})\!\in\![3,5.5]$;
$R\!\in\![0.5,2]\,R_{\mathrm{Jup}}$; $f_{\mathrm{sed}}\!\in\![0.5,8]$; cloud
$\sigma_c\!\in\![1.2,2.5]$; $v_r\!\in\![-100,100]$,
$v\sin i\!\in\![0,30]$\,\si{\km\per\s}; $\sigma_{\mathrm{cal}}\!\in\![0,0.05]$;
SNR conditioning $\in[10,200]$. \textbf{Abundance recipe.} Every element heavier
than helium is first scaled by $10^{[\mathrm{M/H}]}$; oxygen then remains at that
metallicity-scaled solar abundance and carbon is set to C/O times that oxygen
abundance.

\textbf{Recalibration map (\S\ref{sec:results}(c)).} The level-$p$ quantile
becomes the posterior sample-quantile at $R^{-1}(p)$, with $R$ the empirical PIT
CDF on the calibration half. Recalibrating both sides of the ensemble factorial
leaves the marginalized advantage at paired-bootstrap $\Delta$coverage $+0.058$,
$95\%$ $[0.041,0.076]$, at $\Delta$width $-0.048$, $[-0.064,-0.024]$.

\textbf{Regime breakdowns (\S\ref{sec:results}(a,c)).} In quintiles of the deep
temperature knot $T_4$ and of SNR the marginalized ensemble's raw C/O coverage
stays at 0.854--0.915 in every bin, against 0.384--0.468 for the control and
0.667--0.749 for the recalibrated control, so the frozen marginal map does not
fix the per-object conditional bias in any regime. Of the 15 parameters, 6 fail
a Bonferroni-corrected Kolmogorov--Smirnov test of PIT uniformity for the
ensemble (\S\ref{sec:results}(b)).

\textbf{Real sample (\S\ref{sec:results}(g)).} Table~\ref{tab:sample} lists
the 14 products, all \texttt{x1d} files from MAST. Programs 1189 and 2124 were
processed with JWST pipeline versions 2.0.1 and 3.0.0. WISE~0458$+$6434 is a
known binary observed with the IFU, so its spectrum is extracted differently
from the 13 fixed-slit spectra. The preprocessed arrays for all 105 products
of versions 1 and 2 are on the Hugging Face dataset under
\texttt{real\_inputs/}, and \texttt{real\_inputs/objects\_clean14.json} lists
the 14 used here. Line-list error is a shared systematic, so the 14 objects are
not independent realizations of it; a population analysis would need it as a
shared latent. The Luhman\,16AB C/O of \citet{yama2026} sits on a third
convention, a gas-phase CO-to-H$_2$O mixing-ratio quantity where ours is an
elemental input. \citet{hood2024}, by contrast, scale their retrieved oxygen by
$1.3$ for silicate sequestration, so their $0.53$ is commensurate with our bulk
elemental ratio. Each model contributes $2{,}001$ posterior draws per object,
split $667$ per seed for the pooled ensembles. Earlier versions also listed recalibrated
intervals for this object. The frozen maps of \S\ref{sec:results}(c) were not
saved with the release, so this version reports raw intervals only.

\textbf{Deployment stress (\S\ref{sec:results}(f)).} The $1{,}024$ objects are
the first perturbed held-out shards on the Hugging Face dataset after the
finite-pixel filter. Each variant draws one real mask per object with a fixed
seed, and every network in Table~\ref{tab:stress} is scored on the identical
masked spectra, so the columns are matched. Pooled ensembles use $133$ draws per
seed and the control $400$. The un-augmented marginalized ensemble scores $0.894$
unmasked here against $0.901$ in the corresponding version-2 run, within the
binomial standard error of $0.009$. The three mask-augmented networks were
trained from seeds 0, 1 and 2 on the perturbed set. On the same unmasked spectra
their pooled median 90\% C/O interval is $0.209$ of the prior width, against
$0.164$ for the un-augmented ensemble. The script and outputs are in the repository under
\texttt{scripts/v3/} and \texttt{results/v3/}.

\textbf{Broadening stress (\S\ref{sec:limits}).} Every line's Lorentz half-width
in PreMODIT follows
$\mathrm{ngamma}(T,P)=\mathrm{ngamma}_{\mathrm{ref}}[i]\,
(T/T_{\mathrm{ref}})^{-n_{T}[j]}P$ from two lookup grids, so scaling
$\mathrm{ngamma}_{\mathrm{ref}}$ by $1.30$ and shifting $n_{T}$ by $-0.15$ on
the saved operator reproduces a rebuild from a line list whose
$\gamma_{\mathrm{air}}$ or $\alpha_{\mathrm{ref}}$ were scaled and whose
$n_{\mathrm{air}}$ or $n_{T}$ were shifted, grid quantization included. The perturbation covers all
six molecular operators and both alternative line-list caches, so the discrete
axis is untouched. The amplitude is the literature air-versus-H$_2$ separation:
H$_2$-broadened CO$_2$ widths run about $1.2$--$1.4\times$ air and the
temperature exponents differ by $0.1$--$0.3$. The reference set is the
smooth-field-only set of \S\ref{sec:results}(h)
(\texttt{SCULPTOR\_FORCE\_COMBO=0}), which is why its reference C/O coverage is
$0.910$ and reproduces the lower block of Table~\ref{tab:ablation}. Reference
and broadened sets share seed $413612$, 512 spectra, 8 shards and a
bit-identical 71-column parameter table, so the comparison varies one axis; the
survivor counts still differ ($434$ against $444$) because the finite-spectrum
filter is not perfectly matched across the two, the broadened spectra failing it
on a slightly different subset. In the self-normalized input space the
network sees, the per-object rms response over the 433 objects present in both
sets has median $0.0125$, p90 $0.107$ and max $0.59$. Per-parameter posterior width ratios run $0.93$--$1.00$, and clouds, radius,
$\log k_{zz}$ and the radial-velocity nuisance move by $-0.002$ to $+0.004$ in
coverage. Randomizing broadening coefficients within laboratory bounds such as
those of \citet{hosokawa2025} (CH$_4$ in H$_2$/He to $1000$\,K, $5$--$45\%$
below ExoMol at $296$\,K) would anchor that prior in physics, though one
species over one band leaves most of G395H to extrapolation.

\textbf{Evaluation protocols (\S\ref{sec:results}).} The primary table and
SBC/TARP use the $n=2048$ leakage-free factorial split (binomial uncertainty
$\approx0.007$ at nominal $0.90$; TARP 50 reference points); the clean held-out,
seed-saturation and $800$-vs-$3{,}000$-shard pilots use $n=512$; the amplitude
sweep $n\!\approx\!450$; the held-out-cache and deployment-stress sets $n=1024$.
Proportion intervals are Wilson.

\textbf{One-axis ablation (\S\ref{sec:results}(h)).} Two training sets of
$2{,}338$ shards each were generated with one axis active. In
\texttt{discrete\_only} the line-list combination still cycles as
$\mathrm{shard}\bmod4$ while every continuous-field $\sigma$ of
Table~\ref{tab:opamp} is set to zero, so the multiplicative field is the identity
bit for bit; in \texttt{smooth\_only} the field runs at production amplitude and
every shard is pinned to the base combination. One network was trained on each
with the seed, schedule and architecture of marginalized seed 0. Two matched
$512$-object test sets (fresh seed $413612$) were generated the same way and
supply the lower block of Table~\ref{tab:ablation}; $438$ and $434$ spectra
survive the finite-pixel filter.

\textbf{Amplitude sweep and held-out cache (\S\ref{sec:results}(d,e)).} The
sweep uses a fresh seed 314159 and rescales only the continuous field
($n\!\approx\!450$ per cell). At $0.43\times$, $1\times$ (trained), $1.43\times$
and $2.14\times$ the marginalized ensemble covers C/O at 0.94, 0.92, 0.86, 0.75;
a single marginalized seed at 0.90, 0.86, 0.79, 0.70; the control at 0.55, 0.46,
0.40, 0.34. Wilson 95\% intervals on the two misspecified ensemble cells are
$[0.821,0.886]$ at $1.43\times$ and $[0.711,0.790]$ at $2.14\times$. The
held-out-cache model (trained without CH$_4$-ExoMol) reaches C/O 0.821
out-of-cache against 0.790 in-cache, overall 0.833 against 0.832, $n=1024$ per
set.

\textbf{Opacity-perturbation amplitudes.} The continuous multiplicative field is
a 4-mode Fourier expansion in $\log\nu$ with zero-mean Gaussian coefficients
drawn independently per species, per mode and per spectrum;
Table~\ref{tab:opamp} lists the per-species standard deviations of the constant
coefficient ($\sigma_{\mathrm{const}}$) and the four modes
($\sigma_{m},\,m=1$--$4$), in natural-log units. Amplitudes come from the
measured HITEMP-vs-ExoMol inter-list ratio spectra over a $(T,P)$ grid
($T=400$--$1500$\,K, $P=0.1$--$30$\,bar). For CH$_4$ the measured smooth
band-RMS is $\approx3.4\%$ (median over the grid) with a $\approx10.7\%$
residual after the smooth fit, and H$_2$O is smaller; the priors are set well
above those values, deliberately over-dispersed, so the tabulated CH$_4$
$\sigma$ correspond to a pointwise log-opacity standard deviation near $0.4$
rather than to the measured percentages. CO, NH$_3$ and H$_2$S have no second
cache and inherit the mean of the two measured species; CO$_2$ is inflated to
$1.5\times$ the largest measured single-species amplitude as a stand-in for its
dominant line-shape-like residual. CO is perturbed only through the continuous
field. Its inherited $\sigma$ is broader in pointwise amplitude than the
measured H$_2$O and CH$_4$ inter-list differences, but this does not guarantee
coverage of CO-specific structured errors. A second discrete CO cache is
therefore an important extension, given the CO-configuration sensitivity
reported by \citet{yama2026}. The amplitude sweep
(\S\ref{sec:results}) rescales these continuous-field $\sigma$ by a single
factor (reference $=1\times$) while retaining the discrete mixture.

\begin{table}[h]
\centering
\caption{C/O central-interval coverage on the $2{,}048$-object leakage-free
factorial test split (split seed $20260716$), plus the overall (15-parameter
mean) at nominal $0.90$. ``control''/``marg s0'' are single networks;
``ctrl-ens.''/``marg 3-seed ens.'' pool three seeds, so the single-network and
ensemble columns are the two compute-matched comparisons. Binomial SE at nominal
$0.90$ is $\approx0.007$; ``width/prior'' is the median $90\%$ C/O interval
width over the prior width ($1.4$).}
\label{tab:coverage}
\begin{tabular}{lcccc}
\toprule
Nominal (C/O) & control s0 & ctrl-ens. & marg s0 & \textbf{marg 3-seed ens.} \\
\midrule
0.50 & 0.195 & 0.342 & 0.442 & \textbf{0.513} \\
0.68 & 0.284 & 0.448 & 0.599 & \textbf{0.695} \\
0.90 & 0.433 & 0.614 & 0.824 & \textbf{0.894} \\
0.95 & 0.504 & 0.680 & 0.884 & \textbf{0.934} \\
\midrule
Overall @0.90     & 0.747 & 0.845 & 0.831 & \textbf{0.902} \\
Width/prior @0.90 & 0.055 & 0.081 & 0.142 & \textbf{0.168} \\
\bottomrule
\end{tabular}
\end{table}

\begin{table}[h]
\centering
\caption{One-axis ablation, C/O central-interval coverage at nominal $0.90$.
Upper block: three single networks that differ only in which axis their training
data varied, evaluated on the $2{,}048$-object test half of the same
leakage-free perturbed split as Table~\ref{tab:coverage}, raw and after the
frozen recalibration map of \S\ref{sec:results}(c) fit on the matched
$2{,}048$-object calibration half. The three arms are matched at $2{,}338$
training shards, seed 0 and 60 epochs (the production models of
Table~\ref{tab:coverage} use $3{,}000$ shards; the arms need only match each
other), and the marginalized seed-0 row is the Table~\ref{tab:coverage} network
re-drawn in this evaluation, hence $0.825$ here against $0.824$ there. Lower
block: the production fixed-opacity control and the three-seed marginalized
ensemble, neither retrained, on held-out sets perturbed along one axis only
($438$ discrete-only and $434$ smooth-only spectra; binomial SE
$\approx0.014$).}
\label{tab:ablation}
\begin{tabular}{lcc}
\toprule
\multicolumn{3}{l}{\emph{Training axis} ($n=2{,}048$ test)} \\
Trained on & raw & recalibrated \\
\midrule
discrete swap only        & 0.454 & 0.746 \\
smooth field only         & 0.793 & 0.873 \\
both (marg.\ seed 0)      & \textbf{0.825} & \textbf{0.877} \\
\midrule
\multicolumn{3}{l}{\emph{Test axis} (raw, no retraining)} \\
Perturbed spectra & fixed-opacity control & marg.\ 3-seed ens. \\
\midrule
discrete only ($n=438$)   & 0.614 & \textbf{0.963} \\
smooth only ($n=434$)     & 0.472 & \textbf{0.910} \\
\bottomrule
\end{tabular}
\end{table}

\begin{table}[h]
\centering
\caption{The two stress tests, C/O central-interval coverage at nominal $0.90$
unless a row says otherwise. Upper block: the deployment path of
\S\ref{sec:results}(f) on $1{,}024$ perturbed held-out objects (binomial SE
$\approx0.009$), every column scored on the same masked spectra. ``control s0''
is the fixed-opacity control, seed 0; ``marg.'' is the marginalized ensemble
trained without mask augmentation. ``width/prior'' is the median $90\%$ C/O
interval over the prior width ($1.4$); bold marks the recommended deployment, and
the lower block, having no recommended column, is unbolded. Lower block: the
broadening perturbation of \S\ref{sec:limits} applied to the marginalized
three-seed ensemble, which is not retrained ($434$ reference and $444$ broadened
spectra; binomial SE $\approx0.014$).}
\label{tab:stress}
\begin{tabular}{lccc}
\toprule
\multicolumn{4}{l}{\emph{Deployment stress} ($n=1{,}024$, masks of the 14 real spectra)} \\
Preprocessing & control s0 & marg.\ 3-seed ens. & \textbf{mask-aug.\ 3-seed ens.} \\
\midrule
none (unmasked) & 0.451 & 0.894 & \textbf{0.924} \\
real-mask median fill & 0.430 & 0.683 & \textbf{0.928} \\
full path & 0.456 & 0.702 & \textbf{0.918} \\
\midrule
Overall @0.90, full path & 0.643 & 0.800 & \textbf{0.894} \\
Width/prior, unmasked & 0.051 & 0.164 & 0.209 \\
Width/prior, full path & 0.078 & 0.215 & 0.224 \\
\midrule
\multicolumn{4}{l}{\emph{Broadening stress} ($\gamma\times1.30$, $n_{T}-0.15$;
marg.\ 3-seed ens.)} \\
Parameter & reference & broadened & $\Delta$ \\
\midrule
C/O                & 0.910 & 0.901 & $-0.009$ \\
overall (15-param) & 0.906 & 0.858 & $-0.048$ \\
$T_3$              & 0.885 & 0.613 & $-0.272$ \\
$T_2$              & 0.929 & 0.795 & $-0.134$ \\
$T_4$              & 0.906 & 0.822 & $-0.084$ \\
$T_1$              & 0.915 & 0.833 & $-0.081$ \\
$\log g$           & 0.926 & 0.854 & $-0.073$ \\
$[$M/H$]$          & 0.931 & 0.899 & $-0.032$ \\
\bottomrule
\end{tabular}
\end{table}

\begin{table}[h]
\centering
\caption{Per-species continuous-field perturbation amplitudes (standard
deviations of the zero-mean Gaussian Fourier coefficients, natural-log units),
from \texttt{interlist\_ratio\_priors\_v1.json}, released with the code. CO,
NH$_3$, and H$_2$S share the mean-of-measured amplitude; CO$_2$ is the
$1.5\times$-inflated stand-in.}
\label{tab:opamp}
\begin{tabular}{lccccc}
\toprule
Species & $\sigma_{\mathrm{const}}$ & $\sigma_1$ & $\sigma_2$ & $\sigma_3$ & $\sigma_4$ \\
\midrule
CH$_4$          & 0.226 & 0.244 & 0.184 & 0.114 & 0.057 \\
H$_2$O          & 0.083 & 0.107 & 0.072 & 0.043 & 0.016 \\
CO, NH$_3$, H$_2$S & 0.154 & 0.176 & 0.128 & 0.078 & 0.036 \\
CO$_2$          & 0.339 & 0.366 & 0.276 & 0.171 & 0.085 \\
\bottomrule
\end{tabular}
\end{table}

\begin{table}[h]
\centering
\caption{C/O for the T8 benchmark 2MASS J04151954$-$0935066 (SNR 24.5) against
\citet{hood2024} $0.53\pm0.01$, raw posteriors from every network. ``Contains
$0.53$'' lists the levels at which the central interval contains Hood's central
value; the $\pm0.01$ statistical band is narrower than our endpoint precision
and is not resolved. ``marg.'' networks are trained with opacity randomization
and ``mask-aug.'' networks with opacity randomization and real-mask
augmentation (\S\ref{sec:results}(f)). Ensembles pool $667$ draws per seed;
single networks use $2{,}001$. Bold marks the recommended deployment.}
\label{tab:real}
\begin{tabular}{lcccc}
\toprule
Network & C/O & 68\% interval & 90\% interval & contains $0.53$ \\
\midrule
control, seed 0 & 0.642 & $[0.577,0.714]$ & $[0.542,0.766]$ & neither \\
control, seed 1 & 0.670 & $[0.599,0.740]$ & $[0.560,0.781]$ & neither \\
control, seed 2 & 0.704 & $[0.638,0.781]$ & $[0.598,0.846]$ & neither \\
control, 3-seed ens. & 0.673 & $[0.603,0.745]$ & $[0.560,0.799]$ & neither \\
\midrule
marg., seed 0 & 0.660 & $[0.524,0.816]$ & $[0.425,0.925]$ & 68\% \& 90\% \\
marg., seed 1 & 0.918 & $[0.737,1.097]$ & $[0.635,1.193]$ & neither \\
marg., seed 2 & 0.684 & $[0.516,0.852]$ & $[0.428,0.954]$ & 68\% \& 90\% \\
marg., 3-seed ens. & 0.740 & $[0.575,0.958]$ & $[0.470,1.101]$ & 90\% only \\
\midrule
mask-aug., seed 0 & 0.497 & $[0.374,0.710]$ & $[0.282,0.934]$ & 68\% \& 90\% \\
mask-aug., seed 1 & 0.617 & $[0.459,0.850]$ & $[0.391,1.032]$ & 68\% \& 90\% \\
mask-aug., seed 2 & 0.681 & $[0.454,0.860]$ & $[0.361,1.061]$ & 68\% \& 90\% \\
\textbf{mask-aug., 3-seed ens.} & \textbf{0.595} & $[0.430,0.829]$ & $[0.343,1.009]$ & \textbf{68\% \& 90\%} \\
\bottomrule
\end{tabular}
\end{table}

\begin{table}[h]
\centering
\caption{The real sample: one combined level-3 \texttt{x1d} spectrum per object
from MAST, NIRSpec G395H/F290LP, ordered by the median SNR after resampling
onto the model grid. Slit spectra are from the S200A1 fixed slit. ``Pipeline''
is the JWST calibration pipeline version in the file header. C/O is the
mask-augmented three-seed ensemble median with its 68\% interval. Three objects
fall below the training SNR range and are excluded from the summaries of
\S\ref{sec:results}(g). $^{a}$The T8 benchmark 2MASS J04151954$-$0935066 of
\citet{hood2024}, at its WISE-epoch position.}
\label{tab:sample}
\small
\begin{tabular}{llcccc}
\toprule
Object & Program/obs. & Mode & Pipeline & SNR & C/O (68\%) \\
\midrule
WISEA J235402.79$+$024014.1 & 2124/012 & slit & 3.0.0 & 6.3 & $0.28_{-0.09}^{+0.13}$ \\
WISEU J050305.68$-$564834.0 & 2124/052 & slit & 3.0.0 & 8.0 & $0.31_{-0.11}^{+0.20}$ \\
CWISEP J193518.59$-$154620.3 & 2124/051 & slit & 3.0.0 & 8.1 & $0.62_{-0.17}^{+0.22}$ \\
WISE J140518.39$+$553421.3 & 2124/011 & slit & 3.0.0 & 10.1 & $0.49_{-0.20}^{+0.28}$ \\
WISE J014656.66$+$423410.0 & 2124/056 & slit & 3.0.0 & 11.6 & $0.24_{-0.08}^{+0.12}$ \\
WISE J222055.31$-$362817.4 & 2124/010 & slit & 3.0.0 & 11.9 & $0.35_{-0.17}^{+0.25}$ \\
WISEPA J154151.66$-$225025.2 & 2124/005 & slit & 3.0.0 & 12.0 & $0.41_{-0.13}^{+0.20}$ \\
WISEA J082507.37$+$280548.2 & 2124/004 & slit & 3.0.0 & 12.2 & $0.35_{-0.14}^{+0.21}$ \\
WISE J053516.80$-$750024.9 & 2124/003 & slit & 3.0.0 & 12.8 & $0.50_{-0.19}^{+0.29}$ \\
WISEPA J171104.60$+$350036.8 & 2124/007 & slit & 3.0.0 & 17.4 & $0.43_{-0.13}^{+0.17}$ \\
WISEPA J075108.79$-$763449.6 & 2124/008 & slit & 3.0.0 & 19.2 & $0.40_{-0.13}^{+0.24}$ \\
WISE J041521.21$-$093500.6$^{a}$ & 2124/059 & slit & 3.0.0 & 24.5 & $0.59_{-0.17}^{+0.23}$ \\
WISE J182831.08$+$265037.7 & 1189/011 & slit & 2.0.1 & 45.8 & $0.30_{-0.12}^{+0.20}$ \\
WISE J045853.89$+$643452.9 & 1189/004 & IFU & 2.0.1 & 106.7 & $0.30_{-0.10}^{+0.13}$ \\
\bottomrule
\end{tabular}
\end{table}

\textbf{Spectra.} Modeling grid $1850$--$3550$\,\si{\per\cm} (10{,}000
equally-spaced-in-log-wavenumber points), variable-resolution convolution
$R\!\approx\!2700$; inputs self-normalized \emph{per spectrum} (log, then
subtract that spectrum's own mean and divide by its own standard deviation),
applied identically to simulated and observed spectra.

\textbf{Calibration is not information gain.} Per-spectrum normalization removes
the absolute flux scale, so radius, which enters the emission model almost
entirely through that scale, is close to unidentifiable, as is the
calibration-jitter nuisance $\sigma_{\mathrm{cal}}$. Expressed as median $90\%$
interval width over the $90\%$ width of the uniform prior, the marginalized
network returns $0.97$ for radius, $0.99$ for $\sigma_{\mathrm{cal}}$, $0.99$
for $\log k_{zz}$, $1.00$ for $\sigma_c$ and $0.93$ for $f_{\mathrm{sed}}$
(essentially the prior), against $0.16$ for C/O, $0.05$ for $T_3$, $0.13$ for
$v_r$ and $0.46$ for [M/H]. Coverage on a prior-returning parameter is trivially
near-nominal, so those five are calibrated but uninformative and inflate the
``overall'' mean. On the separate trained-amplitude validation cell
($n\approx453$), restricting the mean to the ten parameters below $0.9$ of the
prior barely moves the ensemble ($0.902\to0.904$) but separates the control
further from it ($0.755\to0.691$), so the reported overall figures are if
anything generous to the control. C/O is firmly in the informative group.
\textbf{Embedding net.} 1D CNN (kernel 7) of stacked convolution, ReLU and
max-pool blocks, then adaptive average pool and a linear map to a 128-d context
vector.
\textbf{Flow.} Bounded conditional neural spline flow (zuko NSF), 10 transforms,
hidden features $(256,256)$; SNR appended to the context. Prior support is
enforced by a bijective element-wise logistic map from the unbounded flow
variable onto $[\mathrm{lo},\mathrm{hi}]$ with the change-of-variable Jacobian
carried in the likelihood, so samples lie in the prior box by construction and
the density stays normalized. This is a bounded transform, \emph{not} post-hoc
clipping of out-of-range samples, which would distort the tails that the coverage
statistics measure.
\textbf{Training.} AdamW, learning rate $10^{-4}$, weight decay $10^{-4}$, batch
64, up to 60 epochs, validation split 0.1, best-validation checkpointing (the
lowest-validation-NLL epoch is restored). Mask augmentation applies a mask drawn at
random from the 105-product set of versions 1 and 2 to each training spectrum
with probability 0.9.
\textbf{Data volume.} Shard size 128 spectra; primary models trained on 3{,}000
shards ($384{,}000$ nominal, $\approx$346k after a strict finite-pixel filter
that drops $\sim$10\%); the scaling point uses 800 shards. That filter is
\emph{not} parameter-independent: applying the identical criterion to a
512-spectrum local validation set, the discard rate rises monotonically across
C/O quintiles, $0.04,\,0.09,\,0.19,\,0.19,\,0.20$ from the lowest to the highest.
Simulator failures therefore tilt the effective training prior away from high
C/O relative to the nominal uniform prior. We have not characterized this
dependence on the full perturbed set, nor propagated the implied prior tilt into
the reported coverages; it is a limitation.
\textbf{Compute.} 1M-spectrum generation on a TPU v4-64; NPE training on the pod
\emph{host CPU}. Simulated-set posterior summaries use 400 samples per network;
the real-data application uses 667 per network so that a three-seed ensemble
also totals 2{,}001 draws.
\textbf{Inference latency (measured).} On a single core of an Intel Core
i5-13600K (\texttt{torch.set\_num\_threads(1)}, CPU-only PyTorch 2.7), one
seed's 400-sample posterior for a single G395H spectrum takes $\approx$1.3\,s
(median of 20 reps; the CNN embedding plus one flow evaluation is
$\approx$0.09\,s and flow sampling adds $\approx$3\,ms/sample); the three-seed
ensemble is $\approx$3.9\,s serially on one core, or $\approx$1.3\,s with the
seeds run concurrently. The cost is thus a fixed forward pass with no per-object
optimization, against $K$ classical retrievals each costing CPU-hours. A GPU or
batched evaluation reduces it further; we report the conservative single-core
figure.

\end{document}